# **Efficient Local and Tabu Search Strategies for Large-Scale Quadratic Integer Programming**

Haibo Wang

A.R. Sánchez, Jr. School of Business, Texas A&M International University, Laredo, TX 78045, USA

hwang@tamiu.edu

956-326-2503

Bahram Alidaee*

School of Business Administration, The University of Mississippi, University, MS 38677, USA

balidaee@bus.olemiss.edu

662-715-1614

*Corresponding author

# Efficient Local and Tabu Search Strategies for Large-Scale General Quadratic Integer Programming

**Abstract**

This study investigates the area of general quadratic integer programming (QIP), encompassing both unconstrained (UQIP) and multi-constrained (CQIP) variants. These NP-hard problems have far-reaching applications, yet non-convex cases have received limited attention in the literature. To address this gap, we introduce a closed-form formula for single-variable changes, establishing novel necessary and sufficient conditions for 1-*Opt* local improvement in UQIP and CQIP. We develop a simple local and sophisticated tabu search with an oscillation strategy tailored for large-scale problems. Experimental results on instances with up to 8,000 variables demonstrate the efficiency of these strategies, producing high-quality solutions within a short time.

## 1. Introduction

The formulation of the unconstrained quadratic integer program (UQIP) is as follows:

$$\textbf{(QP0)} \quad \text{Max } f(x) = d^t x + x^t Q x, \text{ s.t. } x_i \in 0,1, \ldots, u_i, \text{ for } i=1, \ldots, n \tag{1}$$

where $x = (x_1, \cdots, x_n)$ is an $n$-dimensional vector of $n$ integer variables, and $x^t$ transpose of $x$; $u_i$ is the upper bound for $x_i$; $d$ is an $n$-dimensional vector of constants, and $d^t$ is the transpose of $d$; and $Q$ is an $n \times n$ upper triangular matrix of a constant. Note that no assumptions are made on the vector $d$ and matrix $Q$, allowing their entries to take on any real number. Due to significant practical relevance, various adaptations of **QP0** have been proposed to tackle this problem, including the addition of side constraints and box constraints (Liu & Gao, 2015; F. Wang, 2021a). One notable variation is Problem (**QP1**), a constrained quadratic integer program (CQIP) with many practical applications.

$$\textbf{(QP1)} \quad \text{Max } f(x) = d^t x + x^t Q x, \text{ s.t. } Ax \le b, x_i \in 0,1, \ldots, u_i, \text{ for } i=1, \ldots, n \tag{2}$$

In Problem (**QP1**), the matrix $A$ is an $m$ by $n$ matrix of non-negative real numbers, while the vector $b$ consists of $m$ positive real numbers. Here, the matrix $A$ and the right-hand side vector $b$ together indicate that the problem is multi-constrained. Both **QP0** and **QP1** problems are NP-hard (nondeterministic polynomial time-hard), making them challenging to solve in a reasonable amount of time. Even determining the local optimality of the quadratic integer problem (QIP) is NP-hard (Chaovalitwongse et al., 2009).

Quadratic integer programs (QIPs) can be categorized into convex and non-convex types, each exhibiting distinct computational complexities.

- Convex QIPs: When the quadratic matrix $Q$ is positive semidefinite, the objective function remains convex, ensuring a well-behaved optimization landscape. This property allows the use of polynomial-time interior-point methods for solving continuous relaxations. Convex QIPs can be efficiently solved using ellipsoid methods or interior-point methods when integrality constraints are relaxed (Xia et al., 2019).
- Non-convex QIPs: If the quadratic matrix $Q$ is indefinite, the problem becomes non-convex, leading to multiple local minima and increased computational complexity. Such cases often require combinatorial search strategies to effectively explore the solution space. For non-convex QIPs, Karush-Kuhn-Tucker (KKT) computations are commonly employed to identify candidate solutions (Beach et al., 2022).

Equation (2) can be interpreted as follows: $j$=1, …, $m$ represents the budget type, and $i$=1, …, $n$ corresponds to the set of agents receiving budgets. No assumptions are made on matrix $Q$ and vector $d$, allowing their entries to take on any real value.

Both problems, **QP0** and **QP1**, with binary variables ($u_i = 1$), have extensive applications and have been thoroughly investigated by numerous researchers (refer to the survey study by Kochenberger et al. 2014). The general integer variable versions of these problems have been applied in various domains, including portfolio optimization, promotion models, capital budgeting, hydrological studies, and the formulation of graph-theoretic problems (Bretthauer & Shetty, 2002; Gupta et al., 1996; Kellerer et al., 2004; Wang & Cao, 2020). The convex formulations of UQIP and CQIP have received significant attention in the literature (Billionnet et al., 2017; Hua et al., 2008; Kozma et al., 2015; Liu & Gao, 2015; Park & Boyd, 2018). Additionally, the literature extensively discusses reformulations of non-convex UQIP and CQIP into binary optimization, along with alternative mixed-integer programs (MIP) (Bergman et al., 2016; Chaovalitwongse et al., 2009; Gümüş & Floudas, 2005; Hadzic & Hooker, 2006; Quadri & Soutil, 2015).

Various efficient algorithms for large-scale binary variables have been developed for both problems, **QP0** and **QP1**, as summarized in Alidaee et al. (2017), Alidaee & Wang, 2017, 2025, Dahmani and Hifi (2019), Fomeni and Letchford (2013), Glover et al. (1998), Pisinger et al. (2007), and Wang and Alidaee (2019a, 2019b, 2023).

However, for the problem with general integer variables, there is a lack of available procedures that can efficiently solve large-scale realistic problems. Most available algorithms are optimal procedures that are only capable of solving small-sized problems, typically with 50-100 variables (Buchheim et al., 2019; Buchheim & Traversi, 2015; Wang, 2021a). Researchers have made progress on special cases of separable and/or convex problems, with some studies testing problems with 36 to 2,000 variables (Billionnet et al., 2017; Fampa & Nieto, 2018; Hua et al., 2008; Quadri & Soutil, 2015; Wang & Cao, 2020). Erenguc and Benson (1991) provided an early optimal solution for general QIP. Later, Gupta et al. (1996) developed an algorithm for ranking solutions, and Sun et al. (2010) introduced a lower bound using convex relaxation. Recent advances include Wang's (2021b) exact solution procedure for non-convex QIP, Nohra et al.'s (2022) convex quadratic relaxation for mixed-integer QIP, and Bonami et al.'s (2022) machine learning-based approach for mixed-integer quadratic programming (MIQP) solvers. Other recent contributions include Pia's (2023) e-approximation solution for indefinite mixed-integer programs, Kushwah and Sharma's (2024) algorithm for multi-objective integer QIP, Shir and Emmerich's (2025) evolutionary algorithm for multi-objective MIQP, and Fali et al.'s (2024) branch-and-cut method for integer indefinite quadratic bilevel maximization problems.

Genuine real-world problems are complex and multifaceted, typically exhibiting non-convex structures that require heuristic solutions (Boros & Hammer, 2003). Surprisingly, the literature on the linear variant of Problem (**QP1**) is remarkably scarce, as evident in recent studies, such as Akçay et al. (2007) and Alidaee et al. (2018), which propose novel approaches for tackling large-scale problems. A common strategy for addressing both Problems, **QP0** and **QP1,** involves reformulating them as binary optimization problems and applying algorithms designed for binary problems, as seen in works by Hadzic and Hooker (2006) and Bergman et al. (2016). However, this transformation introduces an enormous number of binary variables, necessitating the consideration of all variables in the solution process. In this paper, our findings reveal that only three essential numbers are required for consideration.

To design an effective and robust experimental framework in this study, we define clear objectives for quadratic integer programming (QIP) and formulate specific, testable hypotheses on the proposed algorithms. We create diverse testbeds that represent various problem sizes, structures, and complexities to comprehensively assess the performance of our proposed algorithms. We implement rigorous experimental protocols, including multiple runs with different random seeds and systematic hyperparameter tuning, to ensure a fair comparison of algorithms. To directly address solution quality and algorithmic efficiency, we select appropriate performance metrics such as the best-found solution (BFS), relative percentage deviation (RPD), and time to best (TB). We conduct thorough sensitivity analyses to evaluate potential biases and limitations in the experimental design, based on a statistical analysis of the results, including tests for significance and effect sizes. The performance of our algorithms is validated by comparing results with optimal solutions or bounds obtained using the best available exact QP solver (Gurobi 11.0.2). We make our data and solutions, comprising 455 instances, publicly available to enhance reproducibility and provide robust evidence supporting our findings.

The primary objectives of this paper are twofold:

1. To establish new necessary and sufficient conditions for local optimality of any quadratic integer program (QIP) by utilizing a closed-form formula to change a single variable, with applications to mixed-integer and continuous quadratic programs.
2. To develop straightforward 1-*Opt* strategies and meta-heuristics based on a tabu search with an oscillation strategy for addressing problems **QP0** and **QP1**, particularly in the context of very large-scale applications. We conduct extensive computational experiments on large-scale instances and provide a comparison with off-the-shelf software (Gurobi 11.0.2).

Through these contributions, we seek to enhance our understanding and practical applicability in solving complex problems, offering insights and strategies that demonstrate efficacy, particularly in addressing large-scale scenarios.

The remainder of this paper is organized as follows. Section 2 lays the groundwork by introducing the fundamental mathematical concepts underlying unconstrained quadratic integer programs (UQIPs) and constrained quadratic integer programs (CQIPs). We derive necessary and sufficient conditions for local optimality within each problem domain. We then present a straightforward yet exhaustive 1-*Opt* algorithm and a tailored tabu search with an oscillation strategy for each case. Our proposed algorithms are empirically validated in Section 3 through comprehensive computational experiments on problem instances featuring up to 8,000 variables. This section also includes a detailed comparison with the state-of-the-art software Gurobi 11.0.2, highlighting the performance and applicability of our methodologies in addressing complex optimization challenges, and provides in-depth analytic insights and implications, and Section 4 concludes the paper.

## 2. Literature Review and Methodology

### 2.1 QIP Direct Solution Heuristics Versus Encoding for QUBO

When addressing QIP problems, we face a fundamental choice in methodology: do we solve the problem directly using specialized heuristics for integer variables, or do we transform non-binary integer variables into binary representations to fit quadratic unconstrained binary optimization (QUBO) formulations? While the community has largely gravitated toward QUBO due to its use of quantum annealing, a closer examination of direct solutions suggests that they may offer compelling advantages that are often overlooked.

Take C.C. Chang et al. (2020) as a case in point. The authors attempted to map integer linear programming (ILP) problems to QUBO using binary encoding and slack variables, specifically testing this on minimum dominating set (MDS) problems. The math here gets messy; the encoding transformation produces $\sum_{i=1}^{n} \log_2(U_i + 1)$ binary variables for every integer variable with a range $U_i$. For MDS on general graphs, the required qubits scale as $n_V \log_2 n_V$ before we even account for minor embedding overhead, leading to a massive increase in the dimensionality of the QUBO matrix. This introduces "qubit chaining," where multiple physical qubits must represent a single logical variable, significantly complicating the process. The results were somewhat sobering—ground state probability only reached roughly 0.5 to 0.95, depending on the embedding. The authors themselves admit their results are "limited to small problems" and that the algorithm "outperforms random guessing but is limited to small problems" (C.C. Chang et al., 2020, p. 1).

To sidestep this binary encoding issue, C.-Y. Chang et al. (2020) also experimented with Benders' decomposition to solve mixed-integer programming problems iteratively. They decomposed these into binary QUBO subproblems and linear programming (LP) subproblems. However, this creates a different bottleneck: multiple Benders cuts require new penalty weights, leading to what appears to be a parameter explosion. As the iterations accumulate, the algorithm introduces constraints, and each one requires its own penalty weight $w_t$. The authors acknowledge that this is a "subtle" problem, which forces them to use "heuristic ways" to choose $w_t$. Even with careful tuning, simulated annealing—acting as a proxy for quantum hardware—struggles to satisfy constraints as the problem size grows. The constraint violation metric doesn't hit zero; it converges to a residual of about $10^{-3}$ to $10^{-2}$ (C.-Y. Chang et al., 2020, Figure 7). As the authors note (C.-Y. Chang et al., 2020, p. 8), "The main reason that simulated annealing fails to find the optimal solution is the violation of the constraints cuts... Adding a constraint in the form of an extra quadratic penalty function effectively diminishes the weight share of the previous penalty functions." For a standard unit commitment problem with 6 buses and 2-3 time periods, the algorithm needed 104 to 120 iterations to approach optimality. When you consider that classical branch-and-cut methods can solve these instances in seconds, the quantum approach appears to lag.

We see similar friction in Ajagekar et al. (2022), who developed hybrid methods for job-shop and batch scheduling. They used a hybrid decomposition to solve the relaxed mixed-integer linear programming (MILP) (assignment) classically and handled the sequencing via quantum annealing. Although sequencing variables are naturally binary, they still had to convert constraints into QUBO form using penalty-free encoding. It is worth noting that their claim of being "parameter-free" only really refers to avoiding the tuning of slack variable weights, not the complexity of the reformulation itself. For job-shop tasks, the QUBO dimension is $| S_m^n | \ (|S_m^n| - 1)/2$ per machine, which tends to explode as more jobs are added. Consequently, the experiment size had to be capped to fit the capacity of the D-Wave 2000Q. One has to wonder if the comparison to Gurobi fully accounts for the overhead added by the decomposition step itself; in many cases, direct integer programming heuristics would likely solve these instances much faster.

Bernal et al. (2020) offer another perspective with their Graver augmented multi-seed Algorithm (GAMA). They used quantum annealing to compute the Graver basis for augmentation-based integer optimization. Theoretically, this approach avoids encoding the objective in QUBO by using multiple feasible solutions as starting points instead. However, they convert the Graver basis to a QUBO via Lawrence lifting, which creates a significantly larger auxiliary problem. The overhead of solving this QUBO to find test sets seems to outweigh the benefit of the augmentation step for most problems. Furthermore, the computation of the Graver basis still relies on binary encoding with slack variables. They found that "using anywhere between 10-100% of the Graver Basis yielded very similar results" (Bernal et al. 2020, Section 7.5), implying that a large portion of the Graver basis elements are redundant. This suggests that direct algorithms tailored to specific problems may be more efficient than attempting to apply the Graver basis approach.

The studies above highlight five distinct limitations of current QUBO formulations:

1. The overhead from binary transformation appears practically unavoidable.

2. Tuning penalty parameters remains a largely unsolved challenge.
3. The small-sized instances used in experiments often fail to accurately estimate real hardware constraints.
4. We have yet to see concrete evidence of quantum advantages over direct integer programming heuristics.
5. Reliability concerns persist, with direct solutions via specialized MIQP heuristics likely to outperform QUBO-encoding on computational time, solution quality, and scalability.

Motivated by these observations, we propose new heuristics for UQIP and CQIP that aim to bypass these reformulation bottlenecks.

## 2.2 Mathematical Basics

We begin by tackling problem **QP0**, establishing key results that lay the foundation for developing heuristic algorithms. Specifically, we derive a necessary and sufficient condition for local optimality of the *1-Opt* strategy, providing crucial insight for solving this problem. We then generalize these results to encompass problem **QP1**, broadening their applicability. Notably, our findings have far-reaching implications, as they are universally applicable to any type of quadratic program (QP), including mixed-integer and continuous QPs.

Given a solution vector *x*, for each component $x_i$ (*i=1, …, n*), we define the following:

$$M(x_i) = d_i + \sum_{j<i} q_{j,i} x_j + \sum_{j>i} q_{i,j} x_j \tag{3}$$

Throughout the paper, for the simplicity of notations, we interchangeably use *M(i)* with $M(x_i)$. Furthermore, we define:

$y1 = 0, y2 = -M(i)/q_{i,i}, y^* = -M(i)/2q_{i,i}$, for (*i=1, …, n*, and $q_{i,i} \neq 0$).
$y_{max}$ = the closest integer to $y^*$ when $q_{i,i} < 0$ (if $y^*$ is integer, then $y_{max} = y^*$)

Note that the *y2* and $y^*$ values, and thus $y_{max}$, depend on subscripts *i* and *j*. We use the above symbols for simplicity of notation.

**Observation 1:** Given a solution vector $x$, for any component $x_i$ (*i=1, …, n*), we can express *f(x)* as follows, where *H* is a quantity independent of the variable $x_i$. It is only dependent upon $x_k$, (for *k=1, …, n*, and $k \neq i$).

$$f(x) = f_i(x) + H \tag{4}$$
$$f_i(x) = q_{i,i}x_i^2 + x_i M(i)$$

Note that $f_i(x)$ is a concave quadratic when $q_{i,i} < 0$, convex when $q_{i,i} > 0$, and linear when $q_{i,i} = 0$. Furthermore, we defined $y_{max}$ only when $f_i(x)$ is concave.

**2. 3 Unconstrained QIP (UQIP)**

In **Theorem 1** below, we present a simple necessary and sufficient condition for the local optimality of **QP0**. Details of proof are based on Figures 1, 2, and 3.

**Theorem 1.** A solution $x^*$ is locally optimal for 1-*Opt* search for problem **QP0** if and only if each component $x_i^*$, for *i* = 1, …, *n*, satisfies the following condition:

| $q_{i,i}$ | $M(i)$ | $y2$ | $y_{max}$ | $x_i^*$ |
|---|---|---|---|---|
| $< 0$ | $\leq 0$ | $\leq 0$ | $\leq 0$ | $x_i^* = 0$ |
| $< 0$ | $> 0$ | $> 0$ | $> 0$ | $x_i^* = Min\{y_{max}, u_i\}$ |
| $> 0$ | $\geq 0$ | $\leq 0$ | $-$ | $x_i^* = u_i$ |
| $> 0$ | $< 0$ | $> 0$ | $-$ | If $u_i > y2$ then $x_i^* = u_i$, otherwise $x_i^* = 0$ |
| $= 0$ | $< 0$ | $-$ | $-$ | $x_i^* = 0$ |
| $= 0$ | $> 0$ | $-$ | $-$ | $x_i^* = u_i$ |

**Proof:** Note that, in all situations, feasible solutions must satisfy $0 \leq x_i^* \leq u_i$, for *i=1,…,n*. We now consider three situations, shown in Figures 1-3.

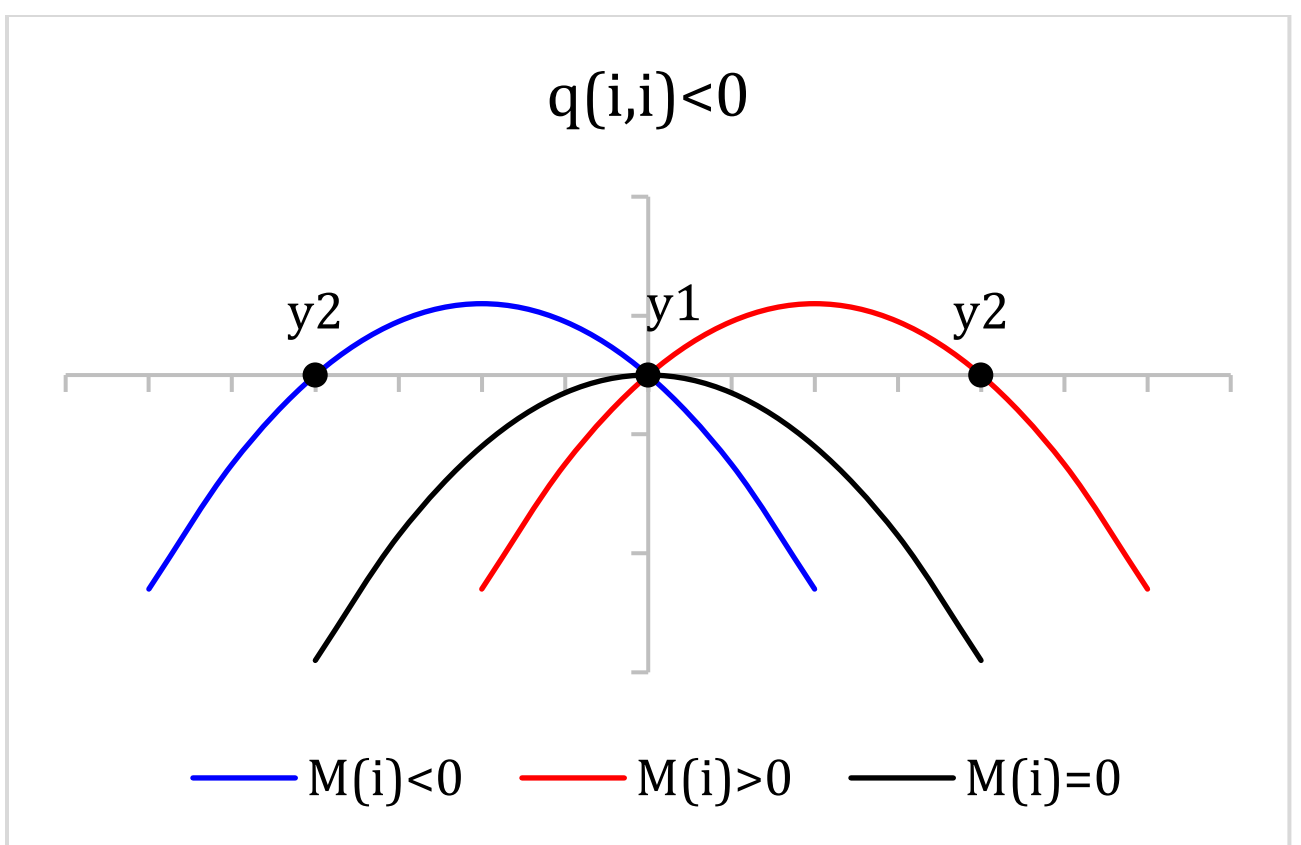

Figure 1. The behavior of the quadratic function $f_i(x)$ when $q_{i,i} < 0$ for different values of *M(i)*

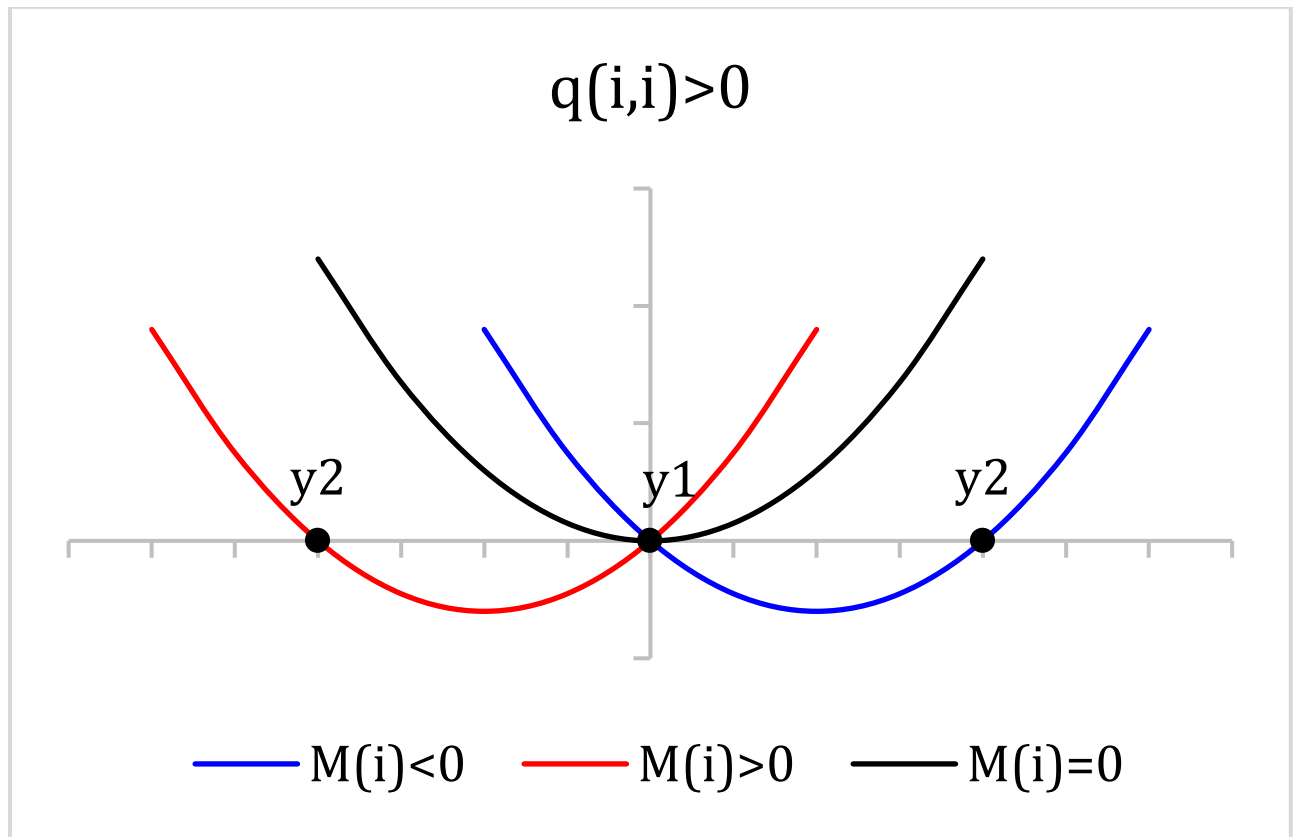

Figure 2. The behavior of the quadratic function $f_i(x)$ when $q_{i,i} > 0$ for different values of *M(i)*

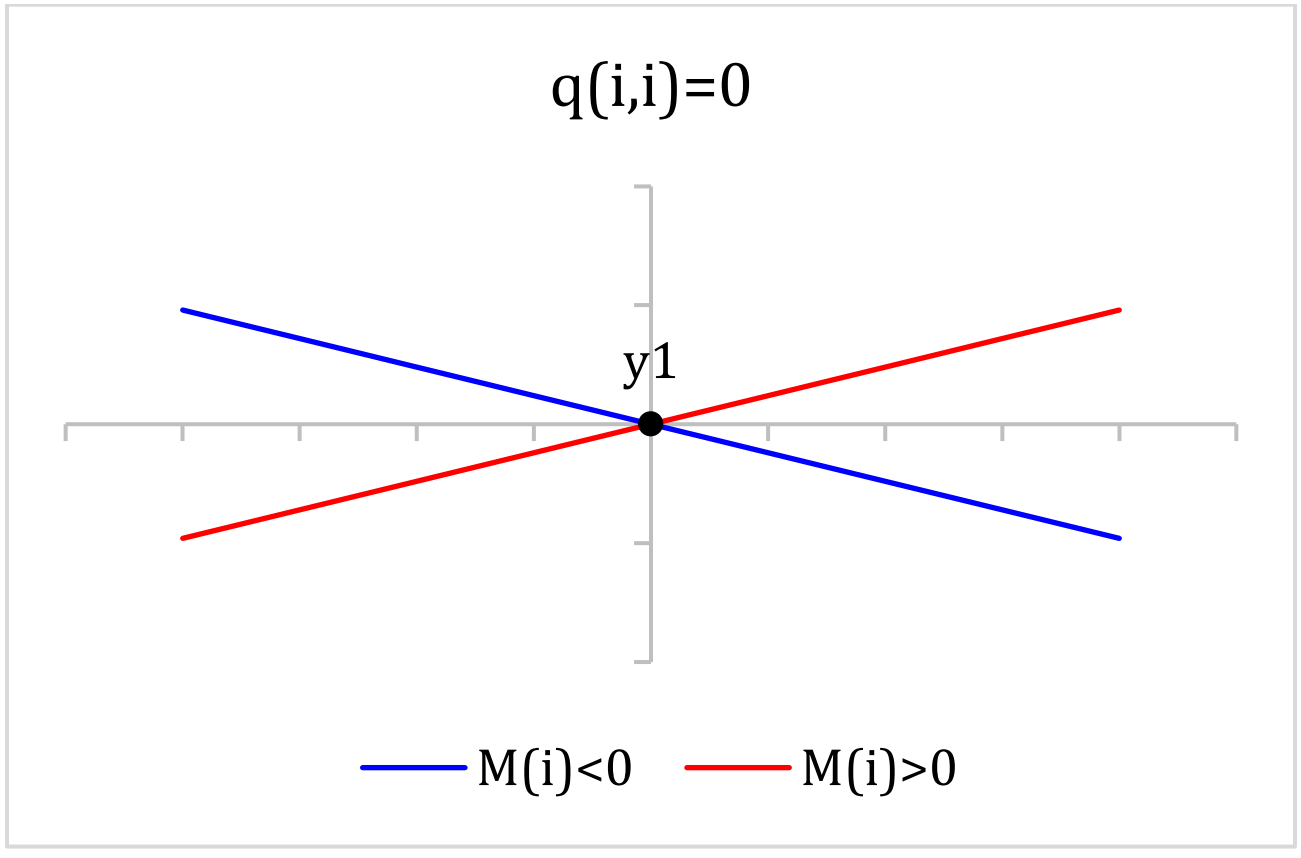

Figure 3. The behavior of the function $f_i(x)$ when $q_{i,i} = 0$ for different values of *M(i)*

If $q_{i,i} < 0$ and $M(i) \leq 0$, all feasible values of $x_i$ makes $f_i(x) \leq 0$ , thus $x_i^* = 0$ is the best choice since $f_i(x^*) = 0$. If $q_{i,i} < 0$ and *M(i)* > 0, values of $x_i$ in the interval $[y1, y2]$ makes $f_i(x) \geq 0$. To satisfy upper bound feasibility, the best choice is thus $x_i^* = Min\{y_{max}, u_i\}$. These situations are shown in Figure 1.

If $q_{i,i} > 0$ and $M(i) \geq 0$, any value of $x_i \geq 0$ makes $f_i(x) \geq 0$. To satisfy upper bound feasibility, the best choice is now $x_i^* = u_i$. If $q_{i,i} > 0$ and *M(i)* < 0, only $x_i > y2$ makes $f_i(x) > 0$. To keep the upper bound feasibility, the best choice of $x_i^*$ must be equal to $u_i$ if $u_i > y2$, otherwise, it is 0. These situations are shown in Figure 2.

If $q_{i,i} = 0$, we have a linear function for $f_i(x)$. Here, we only have two cases. Values of $x_i \geq 0$ for *M(i)* < 0 makes $f_i(x) \leq 0$ . Thus, the best choice is $x_i^* = 0$. However, when *M(i)* > 0, we have $f_i(x) \geq 0$ for all $x_i \geq 0$. To keep the upper bound feasible, the best choice is $x_i^* = u_i$. Figure 3 illustrates these situations.

Given solution *x*, if we modify a component $x_i$ to *y*, the resulting change in the objective function value and the corresponding updates to *M*(*j*) (*j*=1, …, *n*) can be efficiently computed as follows. These updates can be performed in *O(n)* time, ensuring an efficient computation.

$$\begin{aligned} &\text{If } x_i = 0 \text{ and } y>0, \; f(x) = f(x) + \left[q_{i,i}y^2 + M(i)y\right] \\ &\text{If } x_i > 0 \text{ and } y=0, \; f(x) = f(x) - \left[q_{i,i}x_i^2 + M(i)x_i\right] \end{aligned} \tag{5}$$

For:

$j < i, M(j) = M(j) + (y - x_i)q_{j,i}$

$j > i, M(j) = M(j) + (y - x_i)q_{i,j}$

$j = i, \;\; M(j) =$ no change

**Numerical Example for UQIP:** Consider an upper triangular matrix *Q*, vectors *d*, and an upper bound vector *u* as follows.

| | | | | | | | | | | |
|---|---|---|---|---|---|---|---|---|---|---|
| | -6 | -4 | -16 | 18 | 11 | -4 | 14 | 0 | 2 | -4 |
| | | 0 | 8 | 3 | -18 | -1 | -4 | -16 | 9 | 0 |
| | | | 15 | -5 | 13 | 1 | -13 | 11 | -11 | 9 |
| | | | | 14 | 8 | -7 | 13 | -1 | -15 | 1 |
| Q= | | | | | -4 | 5 | 6 | 3 | 0 | 3 |
| | | | | | | -17 | 16 | 3 | 1 | 7 |
| | | | | | | | -17 | 9 | -19 | -18 |
| | | | | | | | | 4 | -16 | 18 |
| | | | | | | | | | 4 | 18 |
| | | | | | | | | | | -19 |
| d(i) | 8 | 1 | 12 | 8 | 8 | 7 | 6 | 5 | 4 | 15 |
| u(i) | 8 | 6 | 7 | 6 | 4 | 8 | 2 | 6 | 3 | 8 |

Let's start with a random solution, $X0$ = (1, 0, 1, 0, 1, 4, 1, 0, 1, 7). Using Theorem 1, we can find M(i), $y2(i)$, $y_{max}$. We randomly chose a component of $X0$ (in this case, $x_4 = 0$, and change according to Theorem 1. It changes to $x_4 = 6$. In the next step, we randomly chose a component of $X1$, in this case, $x_8 = 0$, and according to **Theorem 1**, its value is changed to $x_8 = 6$. The process continues until no more components of $X$ can be changed. This process stops at $X10$=(3, 0, 7, 6, 4, 1, 1, 6, 0, 5) with a local optimal $Z$=2970, which happens to be the optimal value.

| n | X0 | qi,i | M(i) | y2(i) | y(max) | X1 | qi,i | M(i) | y2(i) | y(max) | X2 | qi,i | M(i) | y2(i) | y(max) |
|---|---|---|---|---|---|---|---|---|---|---|---|---|---|---|---|
| **1** | 1 | -6 | -25 | -4.17 | -2 | 1 | -6 | 83 | 13.83 | 7 | 1 | -6 | 83 | 13.83 | 7 |
| **2** | 0 | 0 | -12 | | | 0 | 0 | 6 | | | 0 | 0 | -90 | | |
| **3** | 1 | 15 | 52 | -3.47 | -2 | 1 | 15 | 22 | -1.47 | -1 | 1 | 15 | 88 | -5.87 | -3 |
| **4** | 0 | 14 | 6 | -0.43 | 0 | 6 | 14 | 6 | -0.43 | 0 | 6 | 14 | 0 | 0 | 0 |
| **5** | 1 | -4 | 79 | 19.75 | 10 | 1 | -4 | 127 | 31.75 | 16 | 1 | -4 | 145 | 36.25 | 18 |
| **6** | 4 | -17 | 75 | 4.41 | 2 | 4 | -17 | 33 | 1.94 | 1 | 4 | -17 | 51 | 3 | 2 |
| **7** | 1 | -17 | -68 | -4 | -2 | 1 | -17 | 10 | 0.59 | 0 | 1 | -17 | 64 | 3.76 | 2 |
| **8** | 0 | 4 | 150 | -37.5 | -19 | 0 | 4 | 144 | -36 | -18 | 6 | 4 | 144 | -36 | -18 |
| **9** | 1 | 4 | 106 | -26.5 | -13 | 1 | 4 | 16 | -4 | -2 | 1 | 4 | -80 | 20 | 10 |
| **10** | 7 | -19 | 51 | 2.68 | 1 | 7 | -19 | 57 | 3 | 2 | 7 | -19 | 165 | 8.68 | 4 |
| | Z=-725 | | | | | Z=-185 | | | | | Z=823 | | | | |

| n | X3 | qi,i | M(i) | y2(i) | y(max) | X4 | qi,i | M(i) | y2(i) | y(max) | X5 | qi,i | M(i) | y2(i) | y(max) |
|---|---|---|---|---|---|---|---|---|---|---|---|---|---|---|---|
| **1** | 7 | -6 | 83 | 13.83 | 7 | 7 | -6 | -13 | -2.17 | -1 | 7 | -6 | -1 | -0.17 | 0 |
| **2** | 0 | 0 | -114 | | | 0 | 0 | -66 | | | 0 | 0 | -63 | | |
| **3** | 1 | 15 | -8 | 0.53 | 0 | 7 | 15 | -8 | 0.53 | 0 | 7 | 15 | -11 | 0.73 | 0 |
| **4** | 6 | 14 | 108 | -7.71 | -4 | 6 | 14 | 78 | -5.57 | -3 | 6 | 14 | 99 | -7.07 | -4 |
| **5** | 1 | -4 | 211 | 52.75 | 26 | 1 | -4 | 289 | 72.25 | 36 | 1 | -4 | 274 | 68.5 | 34 |
| **6** | 4 | -17 | 27 | 1.59 | 1 | 4 | -17 | 33 | 1.94 | 1 | 1 | -17 | 33 | 1.94 | 1 |
| **7** | 1 | -17 | 148 | 8.71 | 4 | 1 | -17 | 70 | 4.12 | 2 | 1 | -17 | 22 | 1.29 | 1 |
| **8** | 6 | 4 | 144 | -36 | -18 | 6 | 4 | 210 | -52.5 | -26 | 6 | 4 | 201 | -50.25 | -25 |
| **9** | 1 | 4 | -68 | 17 | 9 | 1 | 4 | -134 | 33.5 | 17 | 1 | 4 | -137 | 34.25 | 17 |
| **10** | 7 | -19 | 141 | 7.42 | 4 | 7 | -19 | 195 | 10.26 | 5 | 7 | -19 | 174 | 9.16 | 5 |
| | Z=1033 | | | | | Z=1705 | | | | | Z=1861 | | | | |

| X6 | qi,i | M(i) | y2(i) | y(max) | X7 | qi,i | M(i) | y2(i) | y(max) | X8 | qi,i | M(i) | y2(i) | y(max) |
|---|---|---|---|---|---|---|---|---|---|---|---|---|---|---|
| 7 | -6 | -3 | -0.5 | 0 | 0 | -6 | -3 | -0.5 | 0 | 0 | -6 | 30 | 5 | 3 |
| 0 | 0 | -72 | | | 0 | 0 | -44 | | | 0 | 0 | -98 | | |
| 7 | 15 | 0 | 0 | 0 | 7 | 15 | 112 | -7.47 | -4 | 7 | 15 | 151 | -10.1 | -5 |
| 6 | 14 | 114 | -8.14 | -4 | 6 | 14 | -12 | 0.86 | 0 | 6 | 14 | 12 | -0.86 | 0 |
| 1 | -4 | 274 | 68.5 | 34 | 1 | -4 | 197 | 49.25 | 25 | 4 | -4 | 197 | 49.25 | 25 |
| 1 | -17 | 32 | 1.88 | 1 | 1 | -17 | 60 | 3.53 | 2 | 1 | -17 | 75 | 4.41 | 2 |
| 1 | -17 | 41 | 2.41 | 1 | 1 | -17 | -57 | -3.35 | -2 | 1 | -17 | -39 | -2.29 | -1 |
| 6 | 4 | 217 | -54.3 | -27 | 6 | 4 | 217 | -54.3 | -27 | 6 | 4 | 226 | -56.5 | -28 |
| 0 | 4 | -137 | 34.25 | 17 | 0 | 4 | -151 | 37.75 | 19 | 0 | 4 | -151 | 37.75 | 19 |
| 7 | -19 | 156 | 8.21 | 4 | 7 | -19 | 184 | 9.68 | 5 | 7 | -19 | 193 | 10.16 | 5 |
| Z=1994 | | | | | Z=2309 | | | | | Z=2840 | | | | |

| X9 | qi,i | M(i) | y2(i) | y(max) | X10 | qi,i | M(i) | y2(i) | y(max) |
|---|---|---|---|---|---|---|---|---|---|
| 0 | -6 | 38 | 6.33 | 3 | 3 | -6 | 38 | 6.33 | 3 |
| 0 | 0 | -98 | | | 0 | 0 | -110 | | |
| 7 | 15 | 133 | -8.87 | -4 | 7 | 15 | 85 | -5.67 | -3 |
| 6 | 14 | 10 | -0.71 | 0 | 6 | 14 | 64 | -4.57 | -2 |
| 4 | -4 | 191 | 47.75 | 24 | 4 | -4 | 224 | 56 | 28 |
| 1 | -17 | 61 | 3.59 | 2 | 1 | -17 | 49 | 2.88 | 1 |
| 1 | -17 | -3 | -0.18 | 0 | 1 | -17 | 39 | 2.29 | 1 |
| 6 | 4 | 190 | -47.5 | -24 | 6 | 4 | 190 | -47.5 | -24 |
| 0 | 4 | -187 | 46.75 | 23 | 0 | 4 | -181 | 45.25 | 23 |
| 5 | -19 | 193 | 10.16 | 5 | 5 | -19 | 181 | 9.53 | 5 |
| Z=2910 | | | | | Z=2970 | | | | |

Building upon Theorem 1, we can efficiently obtain a locally optimal solution using a simple 1-*Opt* local search. Our experimental results confirm that an exhaustive 1-*Opt* local search can rapidly converge to local optimality. To improve solution quality and escape local optimality, we design a hybrid heuristic approach that combines a genetic algorithm with random keys (GARK) and tabu search with an oscillation strategy (TSOS). We present two algorithms for problem **QP0**: **Algorithm 1** integrates an exhaustive 1-*Opt* local search with GARK, and **Algorithm 2** combines GARK with TSOS. The pseudocode of algorithms is presented in the online supplement.

### 2.4 Multi-Constrained QIP (CQIP)

The *quadratic multi-knapsack integer program* (P1) has numerous applications (Kellerer et al., 2004). For binary variables, several algorithms are available for medium- to large-scale problems (Dahmani & Hifi, 2019; Fomeni & Letchford, 2013; Pisinger et al., 2007; Wang et al., 2012). Notably, the objective function of CQIP is identical to that of **QP1**. Therefore, when solving **QP1**, we focus solely on addressing the feasibility issues of the problem. Interestingly, when the objective function only includes the linear term, *dx*, a heuristic capable of handling large-scale problems has been recently developed by Alidaee et al. (2018). Their approach employed surrogate programming within a critical event tabu search, a variant of tabu search with an oscillation strategy, and provided solution procedures for large-scale problems.

Let $a_{j,i}$ denote the *ji*-th element of Matrix *A*. Given a feasible solution *x*, the uncommitted budget from resource *j* (*j*=1, …, *m*), yet available to be allocated to agents *i*=1, …, *n*, denoted by $B_j$ is defined as follows:

$$B_j = b_j - \sum_{i=1}^{n} a_{j,i} x_i, \text{ for } j=1, \ldots, m \tag{6}$$

Also, let $\alpha_i = Min_{1 \le j \le m} \lfloor B_j / q_{j,i} \rfloor$, and $y_i = Min\{u_i, \alpha_i + x_i\}$, for *i*=1, …, *n*.

Given a feasible solution *x*, the value of $\alpha_i$ represents the maximum *additional* amount of resource type *j* that can be allocated to agent *i*, i.e., this amount can be added to the already allocated resource $x_i$ while ignoring the upper-bound constraint. However, the value of $y_i$ represents the maximum total amount of resource *j* that can be allocated to agent *i*, including the existing allocation $x_i$. In that, the upper-bound constraint is accounted for.

Theorem 2 establishes the necessary and sufficient conditions for a local optimum of a solution *x* of problem **QP1**. Details of the proof are based on Figures 1, 2, and 3.

**Theorem 2**. A solution $x^*$ is locally optimal for the 1-*Opt* search for problem **QP1** if and only if each component $x_i^*$, for i = 1, …, *n*, satisfies the following condition:

| $q_{i,i}$ | $M(i)$ | $y2$ | $y_{max}$ | $x_i^*$ |
|---|---|---|---|---|
| < 0 | ≤ 0 | ≤ 0 | ≤ 0 | $x_i^* = 0$ |
| $< 0$ | $> 0$ | $> 0$ | $> 0$ | $\mathrm{x}_i^* = Min\{y_{max}, u_i, \alpha_i + x_i\}$ |
| $> 0$ | $\geq 0$ | $\leq 0$ | $-$ | $\mathrm{x}_i^* = Min\{u_i, \alpha_i + \mathrm{x}_i^*\}$ |
| $> 0$ | $< 0$ | $> 0$ | $-$ | If $u_i > y2$ then $\mathrm{x}_i^* = Min\{u_i, \alpha_i + \mathrm{x}_i^*\}$ otherwise $\mathrm{x}_i^* = 0$ |
| $= 0$ | $< 0$ | $-$ | $-$ | $\mathrm{x}_i^* = 0$ |
| $= 0$ | $> 0$ | $-$ | $-$ | $\mathrm{x}_i^* = Min\{u_i, \alpha_i + \mathrm{x}_i^*\}$ |

**Proof:** Note that, in all situations, feasible solutions must satisfy $0 \leq \mathrm{x}_i^* \leq u_i$, for $i$=1, …, $n$. The proof is similar to that of **Theorem 1**, while accounting for budget constraints. We consider three scenarios, as illustrated in Figures 1-3.

If $q_{i,i} < 0$ and $M(i) \leq 0$, we have a similar situation in **Theorem 1**, thus $\mathrm{x}_i^* = 0$. If $q_{i,i} < 0$ and $M(i) > 0$, values of $x_i$ in the interval [$y1$, $y2$] makes $f_i(x) \geq 0$. To satisfy budget constraints and upper-bound feasibility, using the earlier definition for the maximum amount of budget that can be allocated to component *i,* we must satisfy $\mathrm{x}_i^* = Min\{y_{max}, u_i, \alpha_i + x_i\}$ . Note that, here when $\alpha_i + \mathrm{x}_i^* < Min\{y_{max}, u_i\}$, the best choice is the maximum total amount of resource type *j* that can be allocated to agent *i*, including the existing allocation $\mathrm{x}_i^*$, (illustrated in Figure 1).

If $q_{i,i} > 0$ and $M(i) \geq 0$, similar to the case for **Theorem 1**, any value of $x_i \geq 0$ makes $f_i(x) \geq 0$. To satisfy budget constraints and upper-bound feasibility, the best choice is now $\mathrm{x}_i^* = Min\{u_i, \alpha_i + \mathrm{x}_i^*\}$. If $q_{i,i} > 0$ and $M(i) < 0$, only $x_i > y2$ makes $f_i(x) > 0$. To keep the upper bound and budget constraints feasible, the best choice of $\mathrm{x}_i^*$ must be equal to $\mathrm{x}_i^* = Min\{u_i, \alpha_i + \mathrm{x}_i^*\}$ if $u_i > y2$. Otherwise, it is 0 (refer to Figure 2).

If $q_{i,i} = 0$, again, we have a linear function for $f_i(x)$. Here, we only have two cases. Values of $x_i \geq 0$ for $M(i) < 0$ makes $f_i(x) \leq 0$. In this case, the best choice is $\mathrm{x}_i^* = 0$. However, when $M(i) > 0$, we have $f_i(x) \geq 0$ for all $x_i \geq 0$. To keep budget constraints and upper-bound feasibility, the best choice is $\mathrm{x}_i^* = Min\{u_i, \alpha_i + \mathrm{x}_i^*\}$. Figure 3 illustrates these situations.

Similar to **Algorithms 1** and **2**, we use **Theorem 2** to develop **Algorithms 3** and **4** for CQIP, which are given in the online supplement. Notably, a key aspect of both algorithms is maintaining the feasibility of the solutions throughout their implementation.

The primary objective of this study is to develop a method that efficiently and directly obtains high-quality, locally optimal solutions for extremely large-scale uncapacitated quadratic integer programming (UQIP) and capacitated quadratic integer programming (CQIP) instances, while maintaining a reasonable computational time. In the following section, we comprehensively compare algorithms on massive-scale UQIP and CQIP instances, as well as smaller instances that can be solved to global optimality using Gurobi 11.0.2. It is worth noting that both UQIP and CQIP instances can exhibit convex or non-convex properties, depending on the specific values of the objective coefficients. Interestingly, Gurobi 11.0.2 provides a built-in function to transform non-convex objectives into convex forms, which can be solved using the spatial branching method. Moreover, the default runtime parameters are used throughout the instances.

## 3. Computational Results

This section presents performance evaluations of the proposed algorithms. To assess the effectiveness of our methods, we conducted an extensive series of computational experiments. The algorithms were implemented in Fortran and compiled using the GNU Fortran90 compiler. All experiments were performed on a single core of an Intel Xeon E5-2695 with a 2.10 GHz CPU and 8 GB of memory. To ensure a fair comparison, all computational jobs were submitted through the OpenPBS job management system, allowing for identical CPU and memory usage for each solver and algorithm in the same instance. Table 2 summarizes the test instance generation information used in this study. The notations employed to generate these instances are explained in Table 1.

The datasets for the uncapacitated quadratic integer programming (UQIP) instances, with sizes ranging up to 8,000 variables, are publicly available at https://figshare.com/s/586301dedee29689502c. Similarly, the datasets for the capacitated quadratic integer programming (CQIP) instances, featuring up to 8,000 variables and 4,000 constraints with varying levels of difficulty, are accessible at that same site.

In Table 2, Row *m* and the last two rows are related to constraints in CQIP, and the rest of the information is related to both UQIP and CQIP. Also, note that the last row is associated with the tightness of the right-hand side (RHS) of constraints.

Table 1. Notations used in tables and figures.

| | |
|---|---|
| n | Number of variables |
| m | Number of constraints (a percentage, c, of n) |
| llq (ulq) | The lower (upper) bound of values for the coefficients of the off-diagonal of the quadratic matrix |
| lldiag (uldiag) | The lower (upper) bound of values for the coefficients of the diagonal of the quadratic matrix |
| lld (uld) | The lower (upper) bound of values for the coefficients of linear terms |
| lla (ula) | The lower (upper) bound of values for the left-hand side (LHS) of constraints |
| lls (uls) | The lower (upper) bound of ratios (percentage) of the sum of coefficients of LHS for generating right-hand-side (RHS) constraints |

Table 2. Test instances generation information.

| **Parameter** | **Problem-1(P1)** | **Problem-2(P2)** | **Problem-3(P3)** | **Problem-4(P4)** | **Problem-5 (P5)** |
|---|---|---|---|---|---|
| n | 10-8,000 | 10-8,000 | 10-8,000 | 10-8,000 | 10-8,000 |
| m | c*n | c*n | c*n | c*n | c*n |
| (llq, ulq) | (-20, 20) | (-40, 40) | (-80, 80) | (-160, 160) | (-200, 200) |
| (lldiag, uldiag) | (-20, 20) | (-40, 40) | (-80, 80) | (-160, 160) | (-200, 200) |
| (lld, uld) | (1, 20) | (1, 40) | (1, 80) | (1, 60) | (1, 200) |
| (llx, ulx) | (0, 10) | (0, 20) | (0, 40) | (0, 80) | (0, 100) |
| (lla, ula) | (0, 9) | (0, 19) | (0, 39) | (0, 79) | (0, 99) |
| (lls, uls) | e, d, h | e, d, h | e, d, h | e, d, h | e, d, h |

Note: c = 0.2 and 0.5; e = (0.6, 0.8); d = (0.4, 0.6); and h = (0.2, 0.4).

In tables regarding UQIP, a problem is represented as n*a*-*p*. In that case, *a* is the number of variables, *p* is the Problem-*p*. For example, n1000-3 means a UQIP with 1,000 variables, and the rest of the data is randomly chosen from Column Problem-3 (P3). In tables regarding CQIP, a problem is represented as n*a*m*bg*-*p*. Here, *a* is the number of variables, *b* is the number of constraints, *g* is the level of tightness (from the last row), and *p* is the Problem-*p*. For example, n1000m200e-3 means a CQIP with 1,000 variables, 200 constraints (c = 0.2), RHS randomly chosen in the bracket *e* = (0.6, 0.8), and other features randomly chosen from the brackets in the Column Problem-3 (P3). Tables 3 and 4 present the number of instances of UQIP and CQIP in this study, totaling 455 instances. For UQIP, we have 65 instances with 13 variable sizes and 5 problem types. For CQIP, we have 390 instances with 13 variable sizes, 26 constraint sizes, three levels of difficulty, and 5 problem types. There are 140 very large and challenging instances with variable sizes up to 8,000 and constraint sizes up to 4,000.

Table 3. Instances for UQIP (total of 65 instances).

| | small | | | | | medium | | | | large | | | |
|---|---|---|---|---|---|---|---|---|---|---|---|---|---|
| n | 10 | 20 | 30 | 40 | 50 | 100 | 200 | 400 | 800 | 1,000 | 2,000 | 4,000 | 8,000 |
| P1 | 1 | 1 | 1 | 1 | 1 | 1 | 1 | 1 | 1 | 1 | 1 | 1 | 1 |
| P2 | 1 | 1 | 1 | 1 | 1 | 1 | 1 | 1 | 1 | 1 | 1 | 1 | 1 |
| P3 | 1 | 1 | 1 | 1 | 1 | 1 | 1 | 1 | 1 | 1 | 1 | 1 | 1 |
| P4 | 1 | 1 | 1 | 1 | 1 | 1 | 1 | 1 | 1 | 1 | 1 | 1 | 1 |
| P5 | 1 | 1 | 1 | 1 | 1 | 1 | 1 | 1 | 1 | 1 | 1 | 1 | 1 |

Table 4. Instances for CQIP (total of 390 instances).

| | small | | | | | medium | | | | large | | | |
|---|---|---|---|---|---|---|---|---|---|---|---|---|---|
| n | 10 | 20 | 30 | 40 | 50 | 100 | 200 | 400 | 800 | 1,000 | 2,000 | 4,000 | 8,000 |
| m | 2, 5 | 4, 10 | 6, 15 | 8, 20 | 10, 25 | 20, 50 | 40, 100 | 80, 200 | 160, 400 | 200, 500 | 400, 1,000 | 800, 2,000 | 1,600, 4,000 |
| P1 | 6 | 6 | 6 | 6 | 6 | 6 | 6 | 6 | 6 | 6 | 6 | 6 | 6 |
| P2 | 6 | 6 | 6 | 6 | 6 | 6 | 6 | 6 | 6 | 6 | 6 | 6 | 6 |
| P3 | 6 | 6 | 6 | 6 | 6 | 6 | 6 | 6 | 6 | 6 | 6 | 6 | 6 |
| P4 | 6 | 6 | 6 | 6 | 6 | 6 | 6 | 6 | 6 | 6 | 6 | 6 | 6 |
| P5 | 6 | 6 | 6 | 6 | 6 | 6 | 6 | 6 | 6 | 6 | 6 | 6 | 6 |

The following criteria were used to compare the performance of the algorithms.

- OFV: Best-found objective function value by each algorithm
- BFS: Best solution found among algorithms and solver within CPU time limit (10-run average)
- RPD: The relative percentage deviation (RPD) is employed to assess the relative performance of each algorithm compared to the best-found solution (BFS) on the same instance, calculated as $RPD = 100 \times \frac{BFS - OFV}{BFS}$
- TB[$s$]: Average time (in seconds) to reach the best solution over 10 independent runs

### 3.1 Analysis of Results for UQIP

The preliminary results suggested that **Algorithms 1** and **2** significantly outperform Gurobi 11.0.2 on instances of all sizes. In these problem sizes, *n* ranges from 10 to 50 (small problems), from 100 to 200 (moderate problems), and from 400 to 800 (larger problems), and the results are provided at https://figshare.com/s/586301dedee29689502c. In the preliminary experiment, we allocated 60 seconds for Algorithms 1 and 2, and 7,200 seconds for Gurobi 11.0.2. **Algorithms 1** and **2** obtained the best-known solutions (BSF) for all small-sized problems. However, Gurobi 11.0.2 obtained 20 of the 25 best solutions, missing 5. For 20 moderate-size problems, **Algorithm 2** obtained the best solution in all 20 solved problems, **Algorithm 1** obtained 8 best solutions, and Gurobi 11.0.2 obtained only four best solutions, missing 16. For larger-size problems, **Algorithm 2** also obtained the best solution in all cases; **Algorithm 1** gave 4 best solutions, while Gurobi 11.0.2 obtained only one best solution. Furthermore, consistent with the results of Chaovalitwongse et al. (2009), the preliminary experiments also showed that the best solutions were usually found at the extreme value of $x$. As the problem size increases, more $x_i$ values, ($i$=1, …, $n$), in the best-found solutions occurred at the extreme points, primarily at the upper bound value.

The large UQIP instances in this study are very challenging for the Gurobi 11.0.2 solver. The absence of explicit constraints results in an unbounded feasible region, making it difficult for the Gurobi solver to determine starting points or search directions, which can potentially lead to numerical instability or slow convergence. This lack of bounding constraints also increases the risk of unboundedness, which can be computationally expensive to detect and prove. Furthermore, the absence of constraints limits pre-solve opportunities and the ability to exploit sparsity. Convexity detection becomes more challenging as it depends entirely on the objective function without the aid of a constrained feasible region.

Finally, we applied the algorithms to very large-scale problems, with sizes ranging from 1,000 to 8,000 variables. The results are shown in Table A1 and Figure 4 in the online supplement. For these problems, we allocated 600 seconds for **Algorithms 1** and **2**, and 14,400 seconds for Gurobi 11.0.2.

The results in Table A1 show that **Algorithm 2** obtained the best solutions for all problems. The results of this algorithm were used as the basis for comparison. The average time to reach the best solution for problem sizes of 1,000, 2,000, 4,000, and 8,000 were 43.85, 7.58, 99.56, and 345.73 seconds. The same criteria for **Algorithm 1** were 0.18, 0.91, 3.52, and 19.26 seconds, while those for Gurobi were 2,532.20, 728.60, 2,979.80, and 12,414.80 seconds. However, regarding different problem types, no specific pattern can be discerned.

For each instance, the success rate, defined as the number of runs that reached its best-found solution (BFS) out of 10 runs, is provided in parentheses. For **Algorithm 2** and problem sizes $n$ equal to 1,000 and 2,000, this value was 3.7. For problem sizes 4,000 and 8,000, it was 1. However, for **Algorithm 1**, this value was 1 for all instances. Figure 4 in the online supplement compares RPD results for **Algorithm 1** and Gurobi 11.0.2 for different problem sizes. As the problem size increases, the gap with BFS for Gurobi becomes significantly larger, while for **Algorithm 1**, it remains stable.

Earlier, it was mentioned that preliminary results revealed values of $x_i$ ($i$=1, …, $n$). The BFS usually occurs at extreme points. Figure 5 in the online supplement shows these results for different problem sizes. Interestingly, about 60% of these values occurred at $u_i$ and about 40% at 0. Overall, 99.97% of these values occurred at the extreme points. This is important, especially from the perspective of heuristic development, as it suggests that checking fewer possibilities for each value of $x_i$ ($i$=1, …, $n$) may be sufficient.

### 3.2 Analysis of Results for CQIP

Analysis of experimental results for CQIP should consider the various tightness of constraints, as it affects the problem's difficulty. We considered two possible cases for the number of constraints, namely $c$=0.2 and $c$=0.5, where $m$=$c$*$n$. Similar to the experiments for UQIP, we also conducted preliminary experiments for **Algorithms 3**, **4**, and Gurobi 11.0.2 for CQIP. The results are provided at https://figshare.com/s/afc8cc6d120d1f21a2ce. However, the preliminary results revealed differences with UQIP. Although **Algorithm 4** provided the best results in all cases, with an average of 9.16 seconds, Gurobi software also obtained very good results, often the best or very close to the best solution, with an average of 4,015.712 seconds. **Algorithm 3**, however, was not as effective as expected in the preliminary results, with an average RPD of 12.34 and an average CPU time of 0.42 seconds. However, as the problem size increases, the performance of **Algorithm 3** also improves, as will be seen in the final experiments for large-scale problems. Note that the best solutions for the CQIP problem are almost always at boundary points. Similar to UQIP, we ultimately experimented with algorithms for very large-scale problems, with sizes ranging from 1,000 to 8,000 variables. The results are presented in Tables A2-4 and Figures 6-8 in the online supplement. Here, we allocated 600 seconds for **Algorithms 3** and **4**, and 14,400 seconds for Gurobi 11.0.2.

#### *3.2.1 Regarding Gurobi 11.0.2 Performance*

The average RPD for Type-*h* problems (very tight problems) is approximately 5.5 times larger than the average RPD of Type-*e* and Type-*d* problems. Interestingly, the average time to reach its own best solutions for Type-*h* problems was 57% of the average time for Type-*e* and Type-*d* problems. These differences are more pronounced for cases with a different number of constraints. For $c$=0.2, the average RPD is 6.5 times larger for Type-*h* problems compared to Type-*e* and Type-*d* problems, while the average time to reach its best solutions was 63% of the average time for Type-*e* and Type-*d* problems. For *c*=0.5, the average RPD is 4.8 times larger for Type-*h* problems compared to Type-*e* and Type-*d* problems, while the average time to reach its best solutions was 51% of the average time for Type-*e* and Type-*d* problems. However, other parameters, such as diagonal, off-diagonal, data, and coefficients of *x*, as well as upper bounds for *x*, do not significantly affect the performance of Gurobi. Overall, these results suggest that the tightness of constraints is the most significant factor affecting the performance of the Gurobi solver. The challenges of solving very large CQIP problems with Gurobi's solver are multifaceted and significant, stemming from these problems' inherent complexity of tightness of constraints and the computational resources' practical limitations. As problem sizes include thousands of variables and constraints, memory requirements can become prohibitive, with even moderately large matrices consuming gigabytes of memory, potentially exceeding available system resources. The computational complexity of CQIP problems, which are generally NP-hard, leads to exponential increases in solution time as the problem size grows, making it extremely difficult to solve large instances optimally. Non-convexity in quadratic terms further complicates the solving process, introducing multiple local optima and challenging the solver's ability to find and prove global optimality. While Gurobi is a state-of-the-art solver, it still faces practical limitations when dealing with large problems, often hitting time limits or struggling to progress toward optimality. Numerical stability issues can arise in large-scale CQIP problems, particularly with large coefficients or poor scaling. The pre-solve phase and necessary reformulations for large problems can be time-consuming and memory-intensive. Integer variables introduce another layer of complexity, necessitating branch-and-bound or other discrete optimization techniques that become particularly challenging on a large scale.

### *3.2.2 Regarding* ***Algorithm 3*** *Performance*

*Regarding Problem Type-h (very tight constraints):* The average RPD for **Algorithm 3** across all problems was 7.29, with the average time to reach its best solution being 65.46 seconds. The average time to reach the best solutions for case *c*=0.5 is 8.16 times larger than that for case *c*=0.2, while the average RPD for these two cases is nearly identical.

*Regarding Problem Type-d (medium tight constraints):* The average RPD for **Algorithm 3** across all problems was 1.26, significantly lower than that for Type-*h* problems. However, the average time to reach its best solution was substantially longer, equal to 208.25 seconds, which is 3.2 times longer. The average times to reach the best solutions for the two cases, *c*=0.2 and *c*=0.5, were comparable; the average RPD values for these two cases were also comparable. These findings suggest that the number of constraints did not significantly affect the algorithm's performance.

*Regarding Problem Type-e (loosely tight constraints):* The average RPD for **Algorithm 3** across all problems was 0.76, lower than that for Type-*d* problems. Interestingly, this algorithm once produced the best-found solution (BFS) among the different algorithms. However, the average time to reach its best solution was also longer, equal to 262.58 seconds, which is 1.26 times longer. The average time to reach the best solutions for the two cases, *c*=0.2 and *c*=0.5, were comparable; the average RPD values for these two cases were also comparable. Similar to the case of Type-*d* problems, these findings suggest that the number of constraints did not significantly impact the algorithm's performance.

### *3.2.3 Regarding **Algorithm 4** Performance*

Note that the BFS among all algorithms was consistently provided by **Algorithm 4**. The results of this algorithm served as the basis for comparison with those of other algorithms.

*Regarding Problem Type-h (very tight constraints):* The average time to reach its best-found solution (i.e., BFS) was significantly longer than for **Algorithm 3**, which is 2.3 times larger. The average time to find the best solutions for case *c*=0.5 was slightly longer than for case *c*=0.2 problems.

*Regarding Problem Type-d (medium tight constraints):* The average time to reach its best-found solution (i.e., BFS) was significantly longer than the average time for Type-*e* problems, being 1.54 times larger. The average times to reach the best solutions for cases *c*=0.2 and *c*=0.5 were identical.

*Regarding Problem Type-e (very tight constraints):* The average time to reach its best-found solution (i.e., BFS) was significantly longer than that for Type-*h* problems and comparable to that for Type-*d* problems. The average time to find the best solutions for case *c*=0.5 was marginally longer than for case *c*=0.2 problems.

### *3.2.4 Extreme Values of x in BFS*

Similar to the results of UQIP, over 99.5% of the best-found solutions of CQIP for all problem types also occurred at extreme values for each $x_i$ (*i*=1, …, *n*). Figures 6-8 in the online supplement show these results for different problem sizes and types. However, there are some differences between the results of UQIP and CQIP, as well as among different problem types. As constraints become tighter, more values of $x_i$ occur at the upper bounds. This is evident from Figure 8 in the online supplement, which shows loosely tight problems, where the percentage of extreme values for each $x_i$ is closer. Additionally, in comparison to UQIP, problems of Type-*h* and Type-*d* for CQIP have more values of $x_i$ occurring at the upper bound.

### *3.2.5 Algorithms vs Gurobi Solver*

The RPD between algorithms and the Gurobi solver on large UQIP instances is significantly larger than the RPD between algorithms and the Gurobi solver on large CQIP instances, primarily due to the unbounded feasible region of UQIP. The proposed algorithms are less sensitive to the existence of constraints. UQIP is easier to solve using the proposed algorithms in this study. Additionally, for the Gurobi solver, the choice of underlying solution method becomes crucial for performance, and the specific structure of quadratic terms (sparse vs. dense) can significantly impact solver performance. For large problems, proving optimality or achieving a small optimality gap can be extremely time-consuming, often forcing the solver to settle for good, feasible solutions without global optimality guarantees within the time limit. Gurobi's reported gap value for large instances is a crucial metric that provides insight into the quality of the current best solution relative to the best possible solution. The gap is calculated as the absolute difference between the best bound (*ObjBound*) and the objective value of the best feasible solution (*ObjVal*), divided by the absolute value of *ObjVal*. It's important to note that for very large problems, achieving a small optimality gap or proving optimality can be extremely time-consuming, often requiring users to settle for good, feasible solutions without global optimality guarantees. While Gurobi can handle both constrained and unconstrained QIP problems, the structure provided by constraints often aids in the solution process. For this reason, it is generally recommended to formulate real-world problems with appropriate constraints to improve solver performance and solution reliability when possible.

### *3.2.6 Sensitivity and Reliability Analysis*

This study employs a comprehensive suite of statistical methods to assess the sensitivity and reliability of proposed algorithms. Statistical ranking methods compare multiple algorithms across various problem instances. The study emphasizes the importance of using sufficient runs and problem instances, selecting appropriate significance levels, correcting for multiple comparisons, and reporting both *p*-values and effect sizes to ensure robust and meaningful results. When assessing the performance of various algorithms across different instances or datasets, Friedman's (1937) test is frequently employed as a non-parametric statistical method. It is suitable for comparing the performance of *k* algorithms over *N* datasets, acting as a non-parametric alternative to repeated measures analysis of variance(ANOVA). The test proceeds by ranking the algorithms based on their performance on each dataset. The null hypothesis states that there is no significant difference in the performance ranks of the algorithms. If Friedman's test yields a statistically significant result, it indicates an overall difference among the algorithms, allowing for rejection of the null hypothesis.

Following a significant Friedman test, post-hoc tests are necessary to determine which specific pairs of algorithms exhibit statistically significant differences. Nemenyi's (1963) test is one such post-hoc procedure, used for pairwise comparisons of all algorithms when no control group is designated. It compares the average ranks of the algorithms and controls the familywise error rate, which is the probability of making at least one false discovery across all pairwise comparisons. A significant difference between two algorithms is declared if their average ranks differ by more than a calculated critical difference (CD).

Alternatively, pairwise signed-rank tests, such as the Wilcoxon signed-rank test, can also serve as post-hoc tests after a significant Friedman test (Woolson, 2005). The Wilcoxon signed-rank test is a non-parametric paired difference test that compares two algorithms based on matched data points (i.e., their performance on each dataset). Unlike the simpler sign test, it considers the magnitude and direction of the differences by ranking them. When performing multiple pairwise Wilcoxon tests, applying a family-wise error correction method, such as the Bonferroni method, to adjust the significance level and mitigate the increased risk of Type I errors from multiple comparisons is essential. Reporting results typically involves stating the test statistics, degrees of freedom, and *p*-value for the omnibus test, then detailing which pairwise comparisons were significant according to the chosen post-hoc test and correction method. Tables 5 and 6 present the sensitivity and reliability analysis of the algorithms.

Table 5. Results of Friedman's Test, Nemenyi's Post-Hoc Test, and Pairwise Wilcoxon Test on the OFV of 65 UQIP Instances.

| **Friedman's Test for Gurobi, Algorithm 1, and Algorithm 2** |
| --- |
| Number of complete experiments (blocks/rows): 65<br>Number of algorithms (groups/columns): 3<br>Friedman chi-squared statistic: 74.73<br>*p*-value: 5.91e-17 |
| Note: Friedman's test was conducted on OFV data from 65 instances to compare the performance of Gurobi, **Algorithm 1**, and **Algorithm 2**, yielding a chi-squared statistic of 74.73 (*p*-value = 5.91e-17). This highly significant result indicates consistent and substantial differences in performance ranks among the three algorithms across all instances. Since the *p*-value is far below 0.05, we reject the null hypothesis |

that all algorithms perform equivalently. However, while Friedman's test confirms a significant difference exists, it does not identify which specific algorithm pairs differ or the direction of those differences.

**--- Post-Hoc Analysis ---**

**Nemenyi's Post-Hoc Test Results (pairwise *p*-values):**

| | Gurobi | **Algorithm 1** | **Algorithm 2** |
|---|---|---|---|
| Gurobi | | 2.34E-04 | 3.44E-10 |
| **Algorithm 1** | 2.34E-04 | | 3.33E-02 |
| **Algorithm 2** | 3.44E-10 | 3.33E-02 | |

Note: Nemenyi's post-hoc test followed the significant Friedman test to determine specific algorithm differences. All pairwise comparisons revealed statistically significant differences ($p$-value < 0.05): Gurobi vs. **Algorithm 1** ($p$-value = 2.34E-04), Gurobi vs. **Algorithm 2** ($p$-value = 3.44e-10), and **Algorithm 1** vs. **Algorithm 2** ($p$-value = 3.33E-02). These results confirm that all three approaches differ significantly in OFV performance. Since **Algorithm 2** produced the highest OFV values in this maximization problem, it significantly outperforms both **Algorithm 1** and Gurobi, while **Algorithm 1** also performs significantly differently from Gurobi.

**Pairwise Wilcoxon Test with Bonferroni Correction (*p*-values):**

| | Gurobi | **Algorithm 1** | **Algorithm 2** |
|---|---|---|---|
| Gurobi | | 1.58E-07 | 1.07E-07 |
| **Algorithm 1** | 1.58E-07 | | 1.14E-05 |
| **Algorithm 2** | 1.07E-07 | 1.14E-05 | |

Note: The Pairwise Wilcoxon test with Bonferroni correction was used for further post-hoc analysis, effectively controlling for the Type I error in multiple comparisons. The test produced extremely small $p$-values (Gurobi vs. **Algorithm 1**: 1.58e-07; Gurobi vs. **Algorithm 2**: 1.07e-07; **Algorithm 1** vs. **Algorithm 2**: 1.14e-05), all of which are well below the 0.05 threshold. These results robustly confirm significant differences between every pair of algorithms and reinforce the findings of Nemenyi's test, establishing **Algorithm 2** as the statistically superior approach, followed by **Algorithm 1**, and finally, Gurobi.

Table 6. Results of Friedman's Test, Nemenyi's Post-Hoc Test, and Pairwise Wilcoxon Test on the OFV of 390 CQIP Instances.

**Friedman's Test for Gurobi, Algorithm 3, and Algorithm 4**

Number of complete experiments (blocks/rows): 390

Number of algorithms (groups/columns): 3

Friedman chi-squared statistic: 384.89

*p*-value: 2.65e-84

Note: Friedman's test compared OFV values across 390 instances for three approaches: Gurobi, **Algorithm 3**, and **Algorithm 4**. The analysis yielded an exceptionally large chi-squared statistic of 384.89 with an extremely small *p*-value of 2.65e-84, vastly below the 0.05 significance threshold. This provides overwhelming statistical evidence of significant performance differences among the algorithms. The magnitude of both the test statistics and *p*-value reflects substantial, consistent variations in algorithm performance rankings across this extensive dataset. While this test conclusively establishes that performance differences exist, it cannot identify which specific algorithm pairs differ or the direction of these differences, necessitating subsequent post-hoc analysis.

**--- Post-Hoc Analysis ---**

**Nemenyi's Post-Hoc Test Results (pairwise *p*-values):**

| | Gurobi | **Algorithm 3** | **Algorithm 4** |
|---|---|---|---|
| Gurobi | | 0.23 | 0 |
| **Algorithm 3** | 0.23 | | 0 |
| **Algorithm 4** | 0 | 0 | |

Note: Nemenyi's post-hoc test was conducted following the significant Friedman test to identify specific performance differences between algorithm pairs. Results showed no significant difference between Gurobi and **Algorithm 3** (*p*-value = 0.23 > 0.05), while **Algorithm 4** significantly outperformed both Gurobi and **Algorithm 3** (*p*-value ≈ 0.0 in both comparisons). Since **Algorithm 4** consistently produced the highest OFV values in this maximization problem, these findings statistically confirm that **Algorithm 4** is superior to both approaches. In contrast, Gurobi and **Algorithm 3** perform comparably without statistically significant differences.

**Pairwise Wilcoxon Test with Bonferroni Correction (*p*-values):**

| | Gurobi | **Algorithm 3** | **Algorithm 4** |
|---|---|---|---|
| Gurobi | | 6.12E-01 | 2.66E-50 |
| **Algorithm 3** | 6.12E-01 | | 3.05E-49 |
| **Algorithm 4** | 2.66E-50 | 3.05E-49 | |

Note: The Pairwise Wilcoxon test with Bonferroni correction was used as a conservative post-hoc analysis, closely mirroring Nemenyi's test results. There was no significant difference between Gurobi and **Algorithm 3** (*p*-value = 0.612 > 0.05), while both Gurobi vs. **Algorithm 4** (*p*-value = 2.66e-50) and **Algorithm 3** vs. **Algorithm 4** (*p*-value = 3.05e-49) showed highly significant differences. These results

confirm that **Algorithm 4** is statistically superior to both Gurobi and **Algorithm 3**, which, in turn, do not differ significantly from each other, providing a clear ranking with **Algorithm 4** as the top performer.

### 3.3 Practical Implications

QIP finds applications in various real-world domains, each presenting unique challenges. In control engineering, model predictive control (MPC) for hybrid systems uses QIP to optimize systems with continuous dynamics and logical rules, facing challenges of computation complexity and real-time constraints (Bemporad & Naik, 2018). Portfolio optimization in finance employs QIP for constructing index funds and asset allocation, grappling with semicontinuous constraints and market uncertainties (Mencarelli & D'Ambrosio, 2019). Water resource management utilizes QIP to optimize allocation and distribution, deal with large-scale systems, and address conflicting objectives (Hemker et al., 2008). In the automotive industry, QIP is applied to vehicle decision-making and motion planning for automated driving systems, requiring real-time solving on embedded hardware and handling complex traffic scenarios (Hemker et al., 2008; Quirynen et al., 2025). Communication systems utilize QIP for multiuser detection to estimate information from multiple users in shared channels, thereby addressing scalability issues as the number of users increases (Sankaran & Ephremides, 1998). QIP optimizes the minimum-cost design of reinforced concrete structures, ensuring code compliance while accounting for material and labor costs, and enforcing geometric and dimensional constraints (Guerra et al., 2011). QIP optimizes sustainable reverse logistics for multi-product supply chains under carbon constraints and uncertainty, balancing profit, social impact, and $CO_2$ emission reduction (Al-Refaie & Kokash, 2023).

These applications share common challenges, including computational complexity, scalability, non-convexity, real-time processing requirements, numerical stability, and the need to handle complex constraints (Boros et al., 2011). This ongoing research aims to develop more efficient algorithms that address these challenges and make QIP-based solutions more practical for large-scale, real-world applications.

## 4. Conclusion

This study presents novel closed-form formulas for unconstrained and constrained quadratic integer programming (QIP) problems. We derived several theoretical results for both UQIP and CQIP and subsequently developed an efficient local search strategy to attain local optimality rapidly. Our computational experiments demonstrate the efficacy of these new formulas on UQIP and CQIP instances.

Notably, our results show that for UQIP, the computational time required by the simple 1-*Opt* local search remains consistent across problems of identical size, regardless of the range of objective function coefficients. Furthermore, for CQIP instances, while the computing time of the 1-*Opt* local search remains consistent for problems of the same size and difficulty level, it varies depending on the difficulty level. These findings are particularly significant when implementing local search strategies on very large-scale instances, as they enable the rapid attainment of local optimality. In a future study, we will report on our design of new heuristics that utilize findings on extreme values to accelerate methods for solving large-scale quadratic integer problems.

**Acknowledgements**

The authors are grateful to the editor and referees for their careful reading of this paper and for providing valuable comments. These comments significantly enhanced the paper's presentation. The authors thank Ms. Amy Palacios and Ms. Jessica Cavazos at the Center for the Study of Western Hemispheric Trade for expert editorial assistance, which substantially improved the language and readability of the paper.

## References

Ajagekar, Akshay, Kumail Al Hamoud, and Fengqi You. (2022) "Hybrid classical-quantum optimization techniques for solving mixed-integer programming problems in production scheduling." IEEE Transactions on Quantum Engineering 3, 3102216. doi: 10.1109/TQE.2022.3187367

Akçay, Y., Li, H., & Xu, S. H. (2007). Greedy algorithm for the general multidimensional knapsack problem. *Annals of Operations Research, 150*(1), 17-29. doi:10.1007/s10479-006-0150-4

Al-Refaie, A., & Kokash, T. (2023). Optimization of a sustainable reverse logistics network with multi-objectives under uncertainty. *Journal of Remanufacturing, 13*(1), 1-23. doi:10.1007/s13243-022-00118-5

Alidaee, B., Ramalingam, V. P., Wang, H., & Kethley, B. (2018). Computational experiment of critical event tabu search for the general integer multidimensional knapsack problem. *Annals of Operations Research, 269*(1), 3-19. doi:10.1007/s10479-017-2675-0

Alidaee, B., Sloan, H., & Wang, H. (2017). Simple and fast novel diversification approach for the UBQP based on sequential improvement local search. *Computers & Industrial Engineering, 111*, 164-175. doi:10.1016/j.cie.2017.07.012

Alidaee, B., & Wang, H. (2017). A note on heuristic approach based on UBQP formulation of the maximum diversity problem. *Journal of the Operational Research Society, 68*(1), 102-110. doi:10.1057/s41274-016-0031-4

Alidaee, B., & Wang, H. (2025). Multilevel facility location optimisation: a novel integer programming formulation and approaches to heuristic solutions. International Journal of Production Research, 1–22. doi:10.1080/00207543.2025.2572422

Beach, B., Hildebrand, R., & Huchette, J. (2022). Compact mixed-integer programming formulations in quadratic optimization. *Journal of Global Optimization, 84*(4), 869-912. doi:10.1007/s10898-022-01184-6

Bean, J. C. (1994). Genetic algorithms and random keys for sequencing and optimization. *ORSA journal on computing*, 6(2), 154-160. doi:10.1287/ijoc.6.2.154

Bemporad, A., & Naik, V. V. (2018). A Numerically Robust Mixed-Integer Quadratic Programming Solver for Embedded Hybrid Model Predictive Control. *IFAC-PapersOnLine, 51*(20), 412-417. doi:10.1016/j.ifacol.2018.11.068

Bergman, D., Cire, A. A., Van Hoeve, W.-J., & Hooker, J. (2016). *Decision diagrams for optimization* (Vol. 1). Cham, Switzerland: Springer.

Bernal, David E., Sridhar Tayur, and Davide Venturelli. (2020) "Quantum integer programming (QuIP) 47-779: lecture notes." arXiv preprint arXiv:2012.11382

Billionnet, A., Elloumi, S., Lambert, A., & Wiegele, A. (2017). Using a Conic Bundle Method to Accelerate Both Phases of a Quadratic Convex Reformulation. *INFORMS Journal on Computing, 29*(2), 318-331. doi:10.1287/ijoc.2016.0731

Boros, E., & Hammer, P.L., (2003), Discrete optimization: the state of the art. Amsterdam: Elsevier.

Boros, E., Crama, Y., Werra, D., Hansen, P. and Maffray, F., (2011). The mathematics of Peter L. Hammer (1936–2006): graphs, optimization, and Boolean models. *Annals of Operations Research*, *188*(1), pp.1-18. doi: 10.1007/s10479-011-0913-4

Bonami, P., Lodi, A., & Zarpellon, G. (2022). A Classifier to Decide on the Linearization of Mixed-Integer Quadratic Problems in CPLEX. *Operations Research, 70*(6), 3303-3320. doi:10.1287/opre.2022.2267

Bretthauer, K. M., & Shetty, B. (2002). The nonlinear knapsack problem – algorithms and applications. *European Journal of Operational Research, 138*(3), 459-472. doi:10.1016/S0377-2217(01)00179-5

Buchheim, C., Montenegro, M., & Wiegele, A. (2019). SDP-based branch-and-bound for non-convex quadratic integer optimization. *Journal of Global Optimization, 73*(3), 485-514. doi:10.1007/s10898-018-0717-z

Buchheim, C., & Traversi, E. (2015). On the separation of split inequalities for non-convex quadratic integer programming. *Discrete Optimization, 15*, 1-14. doi:10.1016/j.disopt.2014.08.002

Chang, Chia Cheng, Chih-Chieh Chen, Christopher Koerber, Travis S. Humble, and Jim Ostrowski. (2020) "Integer programming from quantum annealing and open quantum systems." arXiv preprint arXiv:2009.11970

Chang, Chin-Yao, Eric Jones, Yiyun Yao, Peter Graf, and Rishabh Jain. (2020) "On hybrid quantum and classical computing algorithms for mixed-integer programming." arXiv preprint arXiv:2010.07852

Chaovalitwongse, W. A., Androulakis, I. P., & Pardalos, P. M. (2009). Quadratic integer programming: complexity and equivalent forms Quadratic Integer Programming: Complexity and Equivalent Forms. In C. A. Floudas & P. M. Pardalos (Eds.), *Encyclopedia of Optimization* (pp. 3153-3159). Boston, MA: Springer US.

Dahmani, I., & Hifi, M. (2019). A modified descent method-based heuristic for binary quadratic knapsack problems with conflict graphs. *Annals of Operations Research*. 298(1-2), 125-147. doi:10.1007/s10479-019-03290-3

Erenguc, S. S., & Benson, H. P. (1991). An algorithm for indefinite integer quadratic programming. *Computers & Mathematics with Applications, 21*(6), 99-106. doi:10.1016/0898-1221(91)90164-Y

Fali, F., Cherfaoui, Y., & Moulaï, M. (2024). Solving integer indefinite quadratic bilevel programs with multiple objectives at the upper level. *Journal of Applied Mathematics and Computing, 70*(2), 1153-1170. doi:10.1007/s12190-023-01968-3

Fampa, M., & Nieto, F. P. (2018). Extensions on ellipsoid bounds for quadratic integer programming. *Journal of Global Optimization, 71*(3), 457-482. doi:10.1007/s10898-017-0557-2

Fomeni, F. D., & Letchford, A. N. (2013). A Dynamic Programming Heuristic for the Quadratic Knapsack Problem. *INFORMS Journal on Computing, 26*(1), 173-182. doi:10.1287/ijoc.2013.0555

Friedman, M. (1937). The Use of Ranks to Avoid the Assumption of Normality Implicit in the Analysis of Variance. *Journal of the American Statistical Association, 32*(200), 675-701. doi:10.2307/2279372

Glover, F., Kochenberger, G. A., & Alidaee, B. (1998). Adaptive Memory Tabu Search for Binary Quadratic Programs. *Management Science, 44*(3), 336-345. doi:10.1287/mnsc.44.3.336

Guerra, A., Newman, A. M., & Leyffer, S. (2011). Concrete Structure Design using Mixed-Integer Nonlinear Programming with Complementarity Constraints. *SIAM Journal on Optimization, 21*(3), 833-863. doi:10.1137/090778286

Gümüş, Z. H., & Floudas, C. A. (2005). Global optimization of mixed-integer bilevel programming problems. *Computational Management Science, 2*(3), 181-212. doi:10.1007/s10287-005-0025-1

Gupta, R., Bandopadhyaya, L., & Puri, M. C. (1996). Ranking in quadratic integer programming problems. *European Journal of Operational Research, 95*(1), 231-236. doi:10.1016/0377-2217(95)00245-6

Hadzic, T., & Hooker, J. (2006). *Postoptimality analysis for integer programming using binary decision diagrams.* Paper presented at the GICOLAG Workshop (Global Optimization: Integrating Convexity, Optimization, Logic Programming, and Computational Algebraic Geometry), Vienna. Technical report, Carnegie Mellon University.

Hemker, T., Fowler, K. R., Farthing, M. W., & von Stryk, O. (2008). A mixed-integer simulation-based optimization approach with surrogate functions in water resources management. *Optimization and Engineering, 9*(4), 341-360. doi:10.1007/s11081-008-9048-0

Hua, Z., Zhang, B., & Xu, X. (2008). A new variable reduction technique for convex integer quadratic programs. *Applied Mathematical Modelling, 32*(2), 224-231. doi:10.1016/j.apm.2006.11.011

Kellerer, H., Pferschy, U., & Pisinger, D. (2004). Introduction to NP-Completeness of Knapsack Problems. In H. Kellerer, U. Pferschy, & D. Pisinger (Eds.), *Knapsack Problems* (pp. 483-493). Berlin, Heidelberg: Springer Berlin Heidelberg.

Kochenberger, G., Hao, J.-K., Glover, F., Lewis, M., Lü, Z., Wang, H., & Wang, Y. (2014). The unconstrained binary quadratic programming problem: a survey. *Journal of Combinatorial Optimization, 28*(1), 58-81. doi:10.1007/s10878-014-9734-0

Kozma, A., Conte, C., & Diehl, M. (2015). Benchmarking large-scale distributed convex quadratic programming algorithms. *Optimization Methods and Software, 30*(1), 191-214. doi:10.1080/10556788.2014.911298

Kushwah, P., & Sharma, V. (2024). An algorithm to solve multi-objective integer quadratic programming problem. *Annals of Operations Research, 332*(1), 433-459. doi:10.1007/s10479-022-05123-2

Liu, C., & Gao, J. (2015). A polynomial case of convex integer quadratic programming problems with box integer constraints. *Journal of Global Optimization, 62*(4), 661-674. doi:10.1007/s10898-014-0263-2

Mencarelli, L., & D'Ambrosio, C. (2019). Complex portfolio selection via convex mixed-integer quadratic programming: a survey. *International Transactions in Operational Research, 26*(2), 389-414. doi:10.1111/itor.12541

Nemenyi, P. B. (1963). *Distribution-free multiple comparisons.* Doctoral Dissertation, Princeton University, New Jersey, NJ.

Nohra, C. J., Raghunathan, A. U., & Sahinidis, N. V. (2022). SDP-quality bounds via convex quadratic relaxations for global optimization of mixed-integer quadratic programs. *Mathematical Programming, 196*(1), 203-233. doi:10.1007/s10107-021-01680-9

Park, J., & Boyd, S. (2018). A semidefinite programming method for integer convex quadratic minimization. *Optimization Letters, 12*(3), 499-518. doi:10.1007/s11590-017-1132-y

Pia, A. D. (2023). An approximation algorithm for indefinite mixed integer quadratic programming. *Mathematical Programming, 201*(1), 263-293. doi:10.1007/s10107-022-01907-3

Pisinger, W. D., Rasmussen, A. B., & Sandvik, R. (2007). Solution of Large Quadratic Knapsack Problems Through Aggressive Reduction. *INFORMS Journal on Computing, 19*(2), 280-290. doi:10.1287/ijoc.1050.0172

Quadri, D., & Soutil, E. (2015). Reformulation and solution approach for non-separable integer quadratic programs. *Journal of the Operational Research Society, 66*(8), 1270-1280. doi:10.1057/jors.2014.76

Quirynen, R., Safaoui, S., & Cairano, S. D. (2025). Real-Time Mixed-Integer Quadratic Programming for Vehicle Decision-Making and Motion Planning. *IEEE Transactions on Control Systems Technology, 33*(1), 77-91. doi:10.1109/TCST.2024.3449703

Sankaran, C., & Ephremides, A. (1998). Solving a class of optimum multiuser detection problems with polynomial complexity. *IEEE Transactions on Information Theory, 44*(5), 1958-1961. doi:10.1109/18.705573

Shir, O. M., & Emmerich, M. (2025). Multi-Objective Mixed-Integer Quadratic Models: A Study on Mathematical Programming and Evolutionary Computation. *IEEE Transactions on Evolutionary Computation*, 29(3), 661-675. doi:10.1109/TEVC.2024.3374519

Sun, X. L., Li, J. L., & Luo, H. Z. (2010). Convex relaxation and Lagrangian decomposition for indefinite integer quadratic programming. *Optimization, 59*(5), 627-641. doi:10.1080/02331930801987607

Wang, F. (2021a). A successive domain-reduction scheme for linearly constrained quadratic integer programming problems. *Journal of the Operational Research Society*, 72(10), 2317-2330. doi:10.1080/01605682.2020.1784047

Wang, F. (2021b). A successive domain-reduction scheme for linearly constrained quadratic integer programming problems. *Journal of the Operational Research Society, 72*(10), 2317-2330. doi:10.1080/01605682.2020.1784047

Wang, F., & Cao, L. (2020). A new algorithm for quadratic integer programming problems with cardinality constraint. *Japan Journal of Industrial and Applied Mathematics, 37*(2), 449-460. doi:10.1007/s13160-019-00403-0

Wang, H., & Alidaee, B. (2019a). Effective heuristic for large-scale unrelated parallel machines scheduling problems. *Omega, 83*, 261-274. doi:10.1016/j.omega.2018.07.005

Wang, H., & Alidaee, B. (2019b). The multi-floor cross-dock door assignment problem: Rising challenges for the new trend in logistics industry. *Transportation Research Part E: Logistics and Transportation Review, 132*, 30-47. doi:10.1016/j.tre.2019.10.006

Wang, H., & Alidaee, B. (2023). A new hybrid-heuristic for large-scale combinatorial optimization: A case of quadratic assignment problem. *Computers & Industrial Engineering, 179*, 109220. doi: 10.1016/j.cie.2023.109220

Wang, H., Alidaee, B., Ortiz, J., & Wang, W. (2021). The multi-skilled multi-period workforce assignment problem. *International Journal of Production Research, 59*(18), 5477-5494. doi:10.1080/00207543.2020.1783009

Wang, H., Kochenberger, G., & Glover, F. (2012). A computational study on the quadratic knapsack problem with multiple constraints. *Computers & Operations Research, 39*(1), 3-11. doi:10.1016/j.cor.2010.12.017

Woolson, R. F. (2005). Wilcoxon Signed-Rank Test. In *Encyclopedia of Biostatistics*, edited by P. Armitage and T. Colton. John Wiley & Sons, Ltd. doi: 10.1002/0470011815.b2a15177

Xia, W., Vera, J. C., & Zuluaga, L. F. (2019). Globally Solving Nonconvex Quadratic Programs via Linear Integer Programming Techniques. *INFORMS Journal on Computing, 32*(1), 40-56. doi:10.1287/ijoc.2018.0883